\documentclass[a4paper]{jpconf}
\usepackage{graphicx}

\newcommand{\epem}     {\ensuremath{\mathrm{e^+e^-}}}
\newcommand{\roots}    {\ensuremath{\sqrt{s}}}
\newcommand{\rootsp}   {\ensuremath{\sqrt{s'}}}
\newcommand{\znull}    {\ensuremath{\mathrm{Z^0}}}
\newcommand{\mz}       {\ensuremath{m_{\znull}}}
\newcommand{\as}       {\ensuremath{\alpha_{\mathrm{S}}}}
\newcommand{\asmz}     {\ensuremath{\as(\mz)}}
\newcommand{\order}[1] {\mbox{\ensuremath{{\cal O}(#1)}}}
\newcommand{\rfour}    {\mbox{\ensuremath{R_4}}}
\newcommand{\ycut}     {\mbox{\ensuremath{y_{\mathrm{cut}}}}}
\newcommand{\evis}     {\ensuremath{E_{\mathrm{vis.}}}}

\newcommand{\expt}     {\ensuremath{\mathrm{(exp.)}}}
\newcommand{\had}      {\ensuremath{\mathrm{(had.)}}}
\newcommand{\theo}     {\ensuremath{\mathrm{(theo.)}}}
\newcommand{\sigtot}   {\ensuremath{\sigma_{\mathrm{tot.}}}}
\newcommand{\sigl}     {\ensuremath{\sigma_{\mathrm{L}}}}
\newcommand{\rz}       {\ensuremath{R_{\mathrm{Z}}}}
\newcommand{\gammah}   {\ensuremath{\Gamma_{\mathrm{h}}}}
\newcommand{\sigmah}   {\ensuremath{\sigma_{\mathrm{h}}}}
\newcommand{\sigmal}   {\ensuremath{\sigma_{\ell}}}
\newcommand{\rtau}     {\ensuremath{R_{\tau}}}
\newcommand{\ddel}     {\ensuremath{\mathrm{d}}}

\begin{document}

\title{\as\ from LEP}

\author{S Kluth}

\address{Max-Planck-Institut f\"ur Physik, F\"ohringer Ring 6, 
         D-80805 M\"unchen, Germany}

\ead{skluth@mppmu.mpg.de}

\begin{abstract}
Recent results on measurements of the strong coupling \as\ from
LEP are reported.  These include analyses of the 4-jet rate using
the Durham or Cambridge algorithm, of hadronic \znull\ decays
with hard final state photon radiation, of scaling violations of the
fragmentation function, of the longitudinal cross section, of the
\znull\ lineshape and of hadronic $\tau$ lepton decays.
\end{abstract}

\section{Introduction}

Determinations of the strong coupling constant \as\ using data from
the LEP experiments ALEPH, DELPHI, L3 and OPAL are among the most
precise and least ambiguous measurements~\cite{kluth06}.  The initial
state is purely leptonic avoiding initial and final state interference,
the data samples are generally large and the experiments have large
acceptances, good resolutions and low backgrounds resulting in small
experimental corrections.  When the re-analysed data from the PETRA
experiment JADE are included a centre-of-mass (cms) energy range of
more than an order of magnitude is covered by many sets of
measurements.

The LEP program ran from 1989 to 2000 and covered cms energies from
89 to 209~GeV.  The samples of hadronic events in \epem\ annihilation
contain \order{10^{5-6}} events near \mz\ (LEP~1), \order{10^{2-3}}
events at higher energies (LEP~2) and \order{10^{3-4}} events below
\mz\ from JADE between 14 and 44~GeV cms energy.

\section{4-jet rate}

\begin{figure}[h]
\begin{minipage}{19pc}
\includegraphics[width=\textwidth]{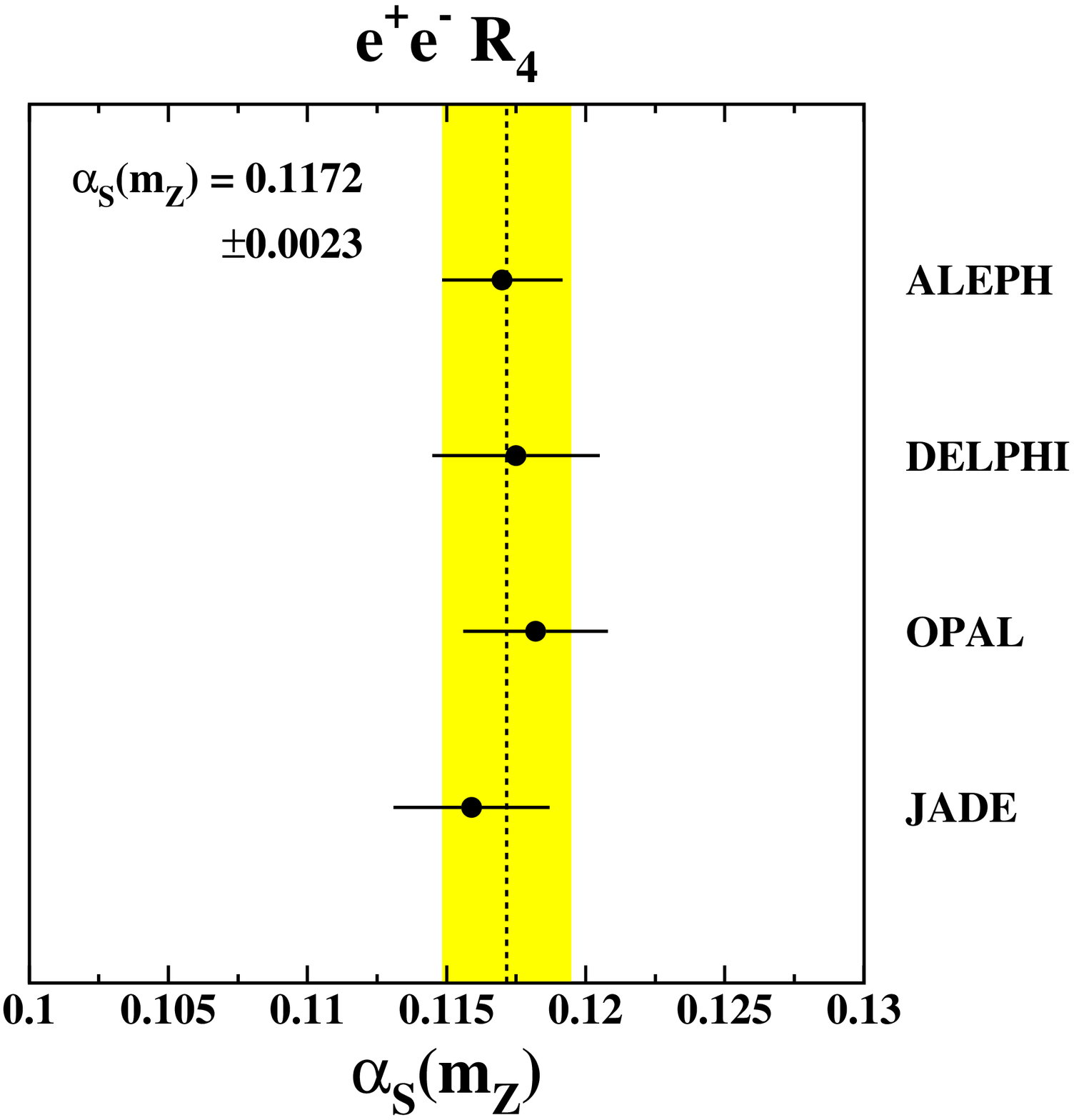}
\caption{Results for \asmz\ from the four-jet rate 
\rfour~\cite{aleph249,delphi335,OPALPR414,jader4}. The dashed vertical
line and shaded band indicate the average value with errors shown on
the figure.}
\label{fig_asr4}
\end{minipage}\hspace{2pc}%
\begin{minipage}{17pc}
\includegraphics[width=\textwidth]{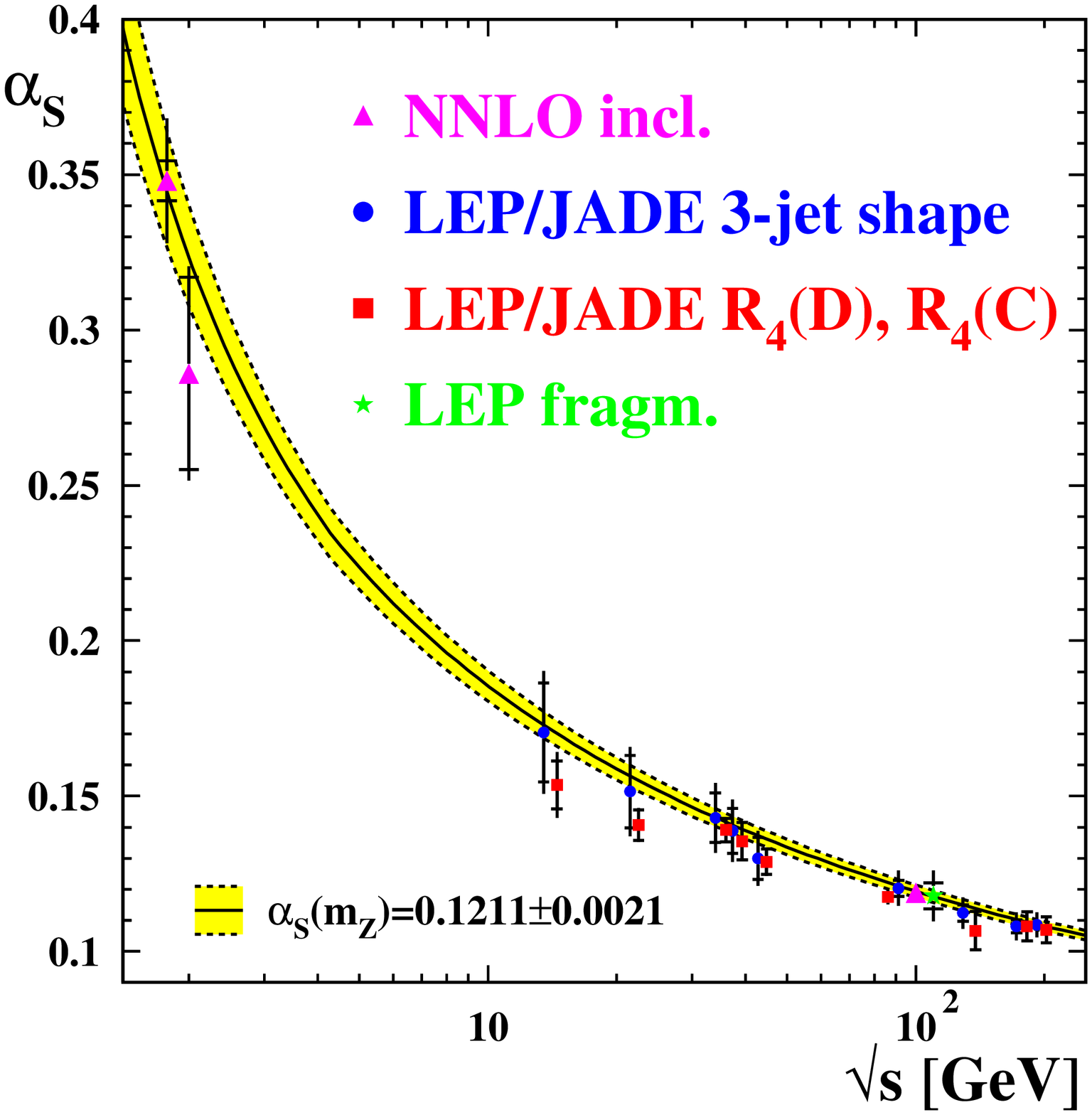}
\caption{Summary of results for \as.  The solid and dashed lines
and the gray band show the the running of \as\ using the average value
of \asmz\ given on the figure.  The label ``NNLO incl.'' refers to
measurements discussed in section~\ref{sec_incl}.  Adapted
from~\cite{kluth06}.}
\label{fig_assum}
\end{minipage}
\end{figure}

In the analyses~\cite{aleph249,delphi335,OPALPR414,jader4} hadronic
final states are clustered using the Durham algorithm, which defines
$y_{ij}=2\min^2(E_i,E_j)(1-\cos\theta_{ij})/\evis^2$ as the phase
space distance between two particles $i,j$ with energies $E_i$ and
$E_j$ with $\evis=\sum_kE_k$.  The pair with the smallest $y_{ij}$ is
combined by adding the four-vectors.  The procedure is repeated until
all $y_{ij}>\ycut$.  The fraction of four-jet events $\rfour(\ycut)$
is studied as a function of \ycut.  DELPHI uses the Cambridge variant
of the Durham algorithm.  Since a four-jet final state corresponds to
at least four partons in QCD the prediction is \order{\as^2}\ in
leading order (LO). The relative error $\Delta\as/\as=\Delta
R_4/(2R_4)$, i.e.\ a precise measurement of \as\ is possible if
experimental uncertainties can be kept under control.  The data,
corrected for experimental effects, are compared with NLO calculations
combined with resummed next-to-leading logarithm (NLLA) terms (DELPHI
use NLO only) including hadronisation corrections derived from
simulation.  The results are summarised in figure~\ref{fig_asr4}.  The
average value $\asmz=0.1172\pm0.0010\expt\pm0.0016\had\pm0.0014\theo$
was found assuming partially correlated experimental uncertainties
while hadronisation and theory uncertainties were evaluated by
repeating the procedure with simultaneously changed input values.
Since the dependency of \as\ on the renormalisation scale (RS) in the
theory is at a minimum in the analyses the conventional estimation of
the theoretical uncertainty by variation of the RS may yield an
underestimate~\cite{OPALPR414,jader4}.

\section{Radiative \znull\ decays}

Hadronic decays of the \znull\ with energetic and isolated photons can
be used to study QCD at lower effective cms energies
\rootsp~\cite{OPALPR425,l3290} assuming that the effects
of wide-angle and hard photon radiation and of QCD processes can be
factorised.  The recent OPAL analysis~\cite{OPALPR425} shows a
successful comparison of event shape distributions using simulated
\znull\ decays with hard and isolated photons and using events
generated at lower cms energy.  The background from events with
isolated neutral hadrons misidentified as photons has to be calculated
from data since the rates of isolated neutral hadrons in the
simulations do not agree with measurements.  The events are binned
according to the photon energy corresponding to
$24\le\rootsp\le78$~GeV.  The event shape analysis uses NLO+NLLA QCD
calculations including hadronisation corrections predicted by
simulations; the fitted theory is found to be consistent with the
data.  Within the uncertainties of the analysis assuming
factorisation of hard and isolated photon radiation and QCD radiation
leads to a good description of the data.

\section{Fragmentation}

In studies of fragmentation in \epem\ annihilation to hadrons
production of charged particles is measured.  The distributions of
$x=2p/\roots$, where $p$ is the particle momentum, are independent of
\roots\ in the static quark-parton model.  Dependence on \roots\ is
referred to as scaling violation and can be described by NLO QCD
calculations.  In~\cite{kluth06} the results from ALEPH, DELPHI and
OPAL are averaged giving $\asmz=0.119\pm0.009$.  The distribution of
angles between charged particles and the beam direction allows to
extract the fraction of events where the intermediate gauge boson
appears with longitudinal polarisation, the so-called longitudinal
cross section \sigl.  Since longitudinal polarisation is only possible
with an additional gluon in the final state \sigl\ can be used to
measure \as.  In~\cite{kluth06} measurements of \sigl\ by ALEPH,
DELPHI and OPAL are averaged with the result
$\sigl/\sigtot=0.056\pm0.002$.  This yields $\asmz=0.117\pm0.008$
using a NLO QCD calculation.  The average for \asmz\ from
fragmentation analyses in \epem\ annihilation from~\cite{kluth06} is
$\asmz=0.118\pm0.004\expt\pm0.001\had\pm0.007\theo$.

\section{\znull\ lineshape and hadronic $\tau$ decays}
\label{sec_incl}

In~\cite{kluth06} the results of the \znull\ lineshape
measurements~\cite{OPALPR412} were used to extract measurements of
\asmz\ from four sensitive observables: the hadronic width \gammah, the
ratio of hadronic and leptonic widths \rz, and the total peak cross
sections for hadron and lepton production \sigmah\ and
\sigmal.  The dependence on the RS~\cite{stenzel05} and the
Higgs mass are studied.  The result based on NNLO QCD calculations is
$\asmz=0.1189\pm0.0027\expt\pm0.0015\theo$.

The analysis of the hadronic branching ratio \rtau\ of the $\tau$
lepton in~\cite{erler03,kluth06} uses NNLO QCD calculations.
Contributions from non-perturbative processes are estimated using the
spectral functions $\ddel\rtau/\ddel s$ for final states with even or
odd numbers of pions and turn out to be consistent with zero for
\rtau.  The result is
$\asmz=0.1221\pm0.0006\expt\pm0.0004\had\pm0.0019\theo$.  A similar
analysis~\cite{davier05} uses partially calculated NNNLO terms in
addition and is consistent.  

In both cases the theoretical uncertainty is determined by varying the
renormalisation scale between 0.5 and 2.0 of the central value such
that the theory errors can be compared directly with other results.

\section{Summary}

The most reliable measurements of \as\ are those based on NNLO
calculations and with small hadronisation uncertainties and thus small
model dependence.  The average of the results from
section~\ref{sec_incl} with a related measurement using the total
cross section for hadron production in \epem\ annihilation at low
\roots\ from~\cite{kluth06} is
$\asmz=0.1211\pm0.0010\expt\pm0.0018\theo$.  All other measurements
discussed here as well as recent determinations of world average
values~\cite{pdg06,kluth06b,bethke06} are consistent with this value.
The results are summarised in figure~\ref{fig_assum} adapted
from~\cite{kluth06}.

Further progress can be expected from the use of NNLO calculations in
the analysis of event shape observables~\cite{gehrmannderidder07,ewng}
and from improved analysis of fragmentation data with NNLO
results~\cite{mitov06}.


\end{document}